\documentclass[apj]{emulateapj}

\usepackage{apjfonts}
\usepackage{amssymb, amsmath}
\newcommand\degree{\degr}
\newcommand\degrees\degree

\DeclareSymbolFont{UPM}{U}{eur}{m}{n}
\DeclareMathSymbol{\umu}{0}{UPM}{"16}
\let\oldumu=\umu
\renewcommand\umu{\ifmmode\oldumu\else\math{\oldumu}\fi}
\newcommand\micro{\umu}
\renewcommand\micron{\micro m}
\newcommand\microns \micron

\renewcommand\arcsec[0]{$^{\prime\prime}$}

\let\oldsim=\sim
\renewcommand\sim{\ifmmode\oldsim\else\math{\oldsim}\fi}
\let\oldpm=\pm
\renewcommand\pm{\ifmmode\oldpm\else\math{\oldpm}\fi}
\newcommand\by{\ifmmode\times\else\math{\times}\fi}
\newcommand\ttt[1]{10\sp{#1}}

\newbox{\wdbox}
\renewcommand\c{\setbox\wdbox=\hbox{,}\hspace{\wd\wdbox}}
\renewcommand\i{\setbox\wdbox=\hbox{i}\hspace{\wd\wdbox}}

\newcount\timect
\newcount\hourct
\newcount\minct
\newcommand\now{\timect=\time \divide\timect by 60
         \hourct=\timect \multiply\hourct by 60
         \minct=\time \advance\minct by -\hourct
         \number\timect:\ifnum \minct < 10 0\fi\number\minct}

\catcode`@=11

\newcommand\comment[1]{}
\newcommand\commenton{\catcode`\%=14}
\newcommand\commentoff{\catcode`\%=12}

\renewcommand\math[1]{$#1$}
\newcommand\mathshifton{\catcode`\$=3}
\newcommand\mathshiftoff{\catcode`\$=12}

\comment{the backslash is necessary}

\comment{alignment tab}

\let\atab=&
\newcommand\atabon{\catcode`\&=4}
\newcommand\ataboff{\catcode`\&=12}

\let\oldmsp=\sp
\let\oldmsb=\sb
\def\sp#1{\ifmmode
           \oldmsp{#1}%
         \else\strut\raise.85ex\hbox{\scriptsize #1}\fi}
\def\sb#1{\ifmmode
           \oldmsb{#1}%
         \else\strut\raise-.54ex\hbox{\scriptsize #1}\fi}
\newbox\@sp
\newbox\@sb
\def\sbp#1#2{\ifmmode%
           \oldmsb{#1}\oldmsp{#2}%
         \else
           \setbox\@sb=\hbox{\sb{#1}}%
           \setbox\@sp=\hbox{\sp{#2}}%
           \rlap{\copy\@sb}\copy\@sp
           \ifdim \wd\@sb >\wd\@sp
             \hskip -\wd\@sp \hskip \wd\@sb
           \fi
        \fi}
\def\msp#1{\ifmmode
           \oldmsp{#1}
         \else \math{\oldmsp{#1}}\fi}
\def\msb#1{\ifmmode
           \oldmsb{#1}
         \else \math{\oldmsb{#1}}\fi}
\def\supon{\catcode`\^=7}
\def\supoff{\catcode`\^=12}
\def\subon{\catcode`\_=8}
\def\suboff{\catcode`\_=12}
\def\supsubon{\supon \subon}
\def\supsuboff{\supoff \suboff}

\newcommand\actcharon{\catcode`\~=13}
\newcommand\actcharoff{\catcode`\~=12}

\newcommand\paramon{\catcode`\#=6}
\newcommand\paramoff{\catcode`\#=12}

\comment{And now to turn us totally on and off...}

\newcommand\reservedcharson{\commenton \mathshifton \atabon \supsubon \actcharon
	\paramon}

\newcommand\reservedcharsoff{\commentoff \mathshiftoff \ataboff
	\supsuboff \actcharoff \paramoff}

\catcode`@=12
\reservedcharsoff

\reservedcharson

\comment{ Must have ONLY ONE of these... trust these macros, they work

}

\newcommand{\squishlist}{
 \begin{list}{$\bullet$}
  { \setlength{\itemsep}{1pt}
     \setlength{\parsep}{0pt}
     \setlength{\topsep}{3pt}
     \setlength{\partopsep}{0pt}
     \setlength{\leftmargin}{2.0em}
     \setlength{\labelwidth}{1.5em}
     \setlength{\labelsep}{0.5em} } }

\newcommand{\squishend}{
  \end{list}  }

\reservedcharson
\actcharon

\usepackage{ulem}
\shorttitle{A {\em Hubble Space Telescope} Search for a Sub-Earth-Sized Exoplanet in the GJ 436 System}
\shortauthors{Stevenson {\em et al.}}

\submitted{}

\begin{document}

\title{A {\em Hubble Space Telescope} Search for a Sub-Earth-Sized Exoplanet in the GJ 436 System}

\author{Kevin B.\ Stevenson\altaffilmark{1,2}}
\author{Jacob L.\ Bean\altaffilmark{1}}
\author{Daniel Fabrycky\altaffilmark{1}}
\author{Laura Kreidberg\altaffilmark{1}}
\affil{\sp{1}Department of Astronomy and Astrophysics, University of Chicago, 5640 S Ellis Ave, Chicago, IL 60637, USA}
\affil{\sp{2}NASA Sagan Fellow}

\email{E-mail: kbs@uchicago.edu}

\begin{abstract}
The detection of small planets orbiting nearby stars is an important step towards the identification of Earth twins.  In previous work using the {\em Spitzer Space Telescope}, we found evidence to support at least one sub-Earth-sized exoplanet orbiting the nearby mid-M dwarf star GJ 436.  As a follow up, here we used the {\em Hubble Space Telescope} to investigate the existence of one of these candidate planets, UCF-1.01, by searching for two transit signals as it passed in front of its host star.  Interpretation of the data hinges critically on correctly modeling and removing the WFC3 instrument systematics from the light curves.  Building on previous {\em HST} work, we demonstrate that WFC3 analyses need to explore the use of a quadratic function to fit a visit-long time-dependent systematic.  This is important for establishing absolute transit and eclipse depths in the white light curves of all transiting systems.  The work presented here exemplifies this point by putatively detecting the primary transit of UCF-1.01 with the use of a linear trend.  However, using a quadratic trend, we achieve a better fit to the white light curves and a reduced transit depth that is inconsistent with previous {\em Spitzer} measurements.  Furthermore, quadratic trends with or without a transit model component produce comparable fits to the available data.  Using extant WFC3 transit light curves for GJ436b, we further validate the quadratic model component by achieving photon-limited model fit residuals and consistent transit depths over multiple epochs.  We conclude that, when we fit for a quadratic trend, our new data contradict the prediction of a sub-Earth-sized planet orbiting GJ 436 with the size, period, and ephemeris posited from the {\em Spitzer} data by a margin of 3.1$\sigma$.
\end{abstract}

%%%%%%%%%%%%%%%%%%%%%%%%%%%%%%%%%%%%%%%%%%%%%%%%%%%%%%%%%%%%%%%%%%%%%%%%%%%%%%%
\section{Introduction}
\label{intro}
%%%%%%%%%%%%%%%%%%%%%%%%%%%%%%%%%%%%%%%%%%%%%%%%%%%%%%%%%%%%%%%%%%%%%%%%%%%%%%%

The search for additional short-period planets in the GJ 436 system began shortly after the transit detection and confirmed eccentric orbit of GJ 436b \citep{Gillon2007b,Deming2007}.  In 2008, \citet{Ribas2008} proposed a $\sim$5-$M\sb{\oplus}$ planet on a 5.2-day orbit.  However, the inferred planet was discredited by orbital-dynamic calculations \citep{Bean2008,Demory2009} and the absence of transit timing variations (TTVs) from additional transit events \citep{Alonso2008,Pont2009,Winn2009}.

NASA's EPOXI mission observed GJ 436 nearly continuously for 22 days.  In analyzing these data, \citet{Ballard2010a} ruled out transiting exoplanets $>$2.0 $R\sb{\oplus}$ outside GJ 436b's 2.64-day orbit (out to a period of 8.5 days) and $>$1.5 $R\sb{\oplus}$ interior to GJ 436b, both with a confidence of 95\%.  Aided by a $\sim$70-hour {\em Spitzer} observation of GJ 436 at 8.0 {\microns}, \citet{Ballard2010b} postulated the presence of a 0.75-$R\sb{\oplus}$ planet with a period of 2.1076 days.  However, they did not detect the predicted transit in an 18-hour follow-up observation with {\em Spitzer} at 4.5 {\microns}.  \citet{Ballard2010b} concluded that the candidate transit signals in the EPOXI and {\em Spitzer} data were the result of correlated noise.

In \citet{Stevenson2012}, we reported on evidence for two sub-Earth-sized planet candidates (UCF-1.01 and UCF-1.02) in the GJ 436 system using the {\em Spitzer Space Telescope}.  
Two fortuitous detections of UCF-1.01 appeared during secondary-eclipse observations of GJ 436b, meant to better constrain the latter's atmospheric composition \citep{Stevenson2010}.  Using these data and a tentative detection at 8.0 {\microns} to estimate its orbital period, we extrapolated UCF-1.01 transit times forward by six months and correctly predicted an event during the next observing window.  With these four 4.5-{\micron} {\em Spitzer} measurements, we constrained the orbital period of UCF-1.01 to be 1.365862 {\pm} 0.000008 days and the transit depth to be 190 {\pm} 25 ppm.  This corresponds to a radius of 0.66 {\pm} 0.04~$R\sb{\oplus}$.  High-precision follow-up observations of such a small target are currently only possible with {\em HST}.

In addition to the planet hypothesis, we examined other sources that could mimic the observed periodic decrease in flux.  GJ 436's large proper motion (across the sky) enabled us to eliminate astrophysical false positives such as an eclipsing-binary star system with a confidence of $>5\sigma$.  We also considered the hypothesis that the observed variations are the result of instrumental or stellar instabilities, but concluded that the single-planet hypothesis is 72 times more likely.

In Section \ref{sec:obs}, we discuss the acquisition and reduction of new {\em Hubble Space Telescope (HST)} data intended to investigate the presence of a sub-Earth-sized exoplanet in the GJ~436 system.  We also describe our light-curve model fits and present additional evidence for a non-linear baseline systematic \citep{Stevenson2014a}.  In Section \ref{sec:results}, we discuss our results and place limits on the size of a detectable planet.

%%%%%%%%%%%%%%%%%%%%%%%%%%%%%%%%%%%%%%%%%%%%%%%%%%%%%%%%%%%%%%%%%%%%%%%%%%%%%%%
\section{OBSERVATIONS AND DATA ANALYSIS}
\label{sec:obs}
%%%%%%%%%%%%%%%%%%%%%%%%%%%%%%%%%%%%%%%%%%%%%%%%%%%%%%%%%%%%%%%%%%%%%%%%%%%%%%%

\subsection{Planning}
\label{sec:ttvs}

Prior to scheduling the {\em HST} observations, we explored the effects of transit timing variations (TTVs) for UCF-1.01 due to gravitational interactions with its Neptune-sized planetary companion GJ 436b.  We considered four possible scenarios that included/excluded the 8.0 {\micron} tentative detection of UCF-1.01 and used the early/late transit time from the bimodal distribution in the 18-hour 4.5 {\micron} observation \citep{Stevenson2012}.  Our Markov-chain Monte Carlo (MCMC) simulation also used eight published \citep{Knutson2011} and two unpublished (J. Winn, M. Holman, private communication) transit times for GJ 436b.  Within our {\em HST} observing window, we found that three of the four scenarios predict transit times that are self-consistent to within two minutes, while the fourth scenario deviates by up to seven minutes.  In comparison, propagating the uncertainty in our best-fit orbital period forward by 2.5 years results in a transit time uncertainty of {\pm}8 minutes.  Both values are smaller than the $\sim$20 minute {\em HST} observation start window; therefore, we concluded that we can safely schedule the observations without undue risk of missing the transit during Earth occultation.

All of the scenarios predict TTVs with an amplitude of 6.5 minutes (13 minutes peak-to-peak) and a period of 41.1 days.  The TTV simulations also provide a more likely estimate of UCF-1.01's average orbital period (1.36588145 days).  We adopted this value when scheduling our observations.

\subsection{Observations and Reduction}

In seeking to detect the primary transit of UCF-1.01, {\em HST} observed the GJ 436 system on two occasions (2013 December 17 and 2014 February 14) using the Wide Field Camera 3 (WFC3) instrument with its G141 grism and 256$\times$256 subarray mode (GO Program 13338, PI: Kevin Stevenson).  Each visit consisted of four {\em HST} orbits, with the predicted transit time occurring during the third orbit.  We employed the spatial-scan mode perpendicular to the dispersion direction at a rate of 1.2{\arcsec}s\sp{-1} (150 pixels in height) that alternated between the forward and reverse directions.  Dispersing light from the $H = 6.3$ magnitude star in both the dispersion and spatial directions allowed us to collect a large number of photons without saturating the detector and, thus, achieve the necessary precision to detect a transiting sub-Earth-sized exoplanet.  The median peak pixel fluence over all frames was 31,600 electrons, which is well below the suggested maximum of 40,000 electrons.  To maximize observing efficiency, we forced a buffer dump at the end of each {\em HST} orbit.  Using ``up-the-ramp'' sampling, each 15 second exposure encompassed three non-destructive reads.  In total, we acquired 176 spectra per visit.

Using the reduction, extraction, and calibration steps described by \citet{Stevenson2014a}, we generate band-integrated (white) light curves spanning 1.125 -- 1.65 {\microns}.  We determine that the scan direction is well aligned with the pixel columns; therefore, for the purpose of generating white light curves, we elect not to correct for the spectrum tilt.  We find an optimal extraction box height of 160 pixels per non-destructive read, as defined by the lowest standard deviation of the normalized residuals (SDNR).  For these data, the photon-limited precision is 67~ppm per point and 10~ppm per {\em HST} orbit.  The latter number implies that for an $H = 6.3$ magnitude object, we can achieve a white light curve precision of 10~ppm on the transit or eclipse depth of a planet whose duration spans at least 50 minutes.

\subsection{Light-Curve Systematics and Fits}

For all transiting exoplanet analyses, it is important to accurately model the systematics when fitting the white light curves.  This is because a white light curve model determines the absolute transit or eclipse depth, which helps to establish the presence of scattering or determine an accurate brightness temperature, respectively.  {\em HST}/WFC3 light curves exhibit several well-characterized systematics, including batch-, orbit-, and visit-long effects, that can sometimes be larger than the differential signals we are trying to detect.  Here we summarize previous work relating to these systematics and discuss how me mitigate their effects on our search for a transit signal in the band-integrated data.  

Using observations of GJ~1214 acquired in WFC3's stare mode, \citet{Berta2012} report ``ramp''-like features within each batch of exposures between buffer downloads.  They link the strength of this systematic to the illumination that each pixel receives, showing that fluence levels as low as 30,000 to 40,000 electrons are correlated with exposure number within each batch.  The WASP-12 light curves first presented by \citet{Swain2012} corroborate these findings as they achieve a maximum fluence of <40,000 electrons and do not exhibit measurable ramp-like features within each batch.  Similarly, after examining eight WFC3 light curves in stare mode, \citet{Wilkins2014} suggest that a dependence becomes apparent at a fluence level of about 30,000 electrons.  With the aim of eliminating the effects of this systematic, we acquired our data using WFC3's spatial scan mode and limited our peak pixel fluence to <32,000 electrons.

Despite using WFC3's spatial scan mode and achieving peak pixel counts near 23,000, \citet{Kreidberg2014} detect a systematic similar in shape to the ramp-like feature discussed above.  However, this systematic typically exhibits a shallower curve, taking up to 45 minutes to reach its steady state, and repeats every {\em HST} orbit.  The magnitude of the ramp is most significant during an observation's first orbit, but appears repeatable over all remaining orbits within the visit.  Both this ramp and the effect observed by \citet{Berta2012} in the stare mode data are likely related to the same physical process that gives rise to detector persistence. This process is suspected to fill charge traps caused by impurities in the HgCdTe photodiodes in the WFC3 IR detector \citep{Long2012}.

\begin{figure}[tb]
\centering
\includegraphics[width=0.9\linewidth,clip]{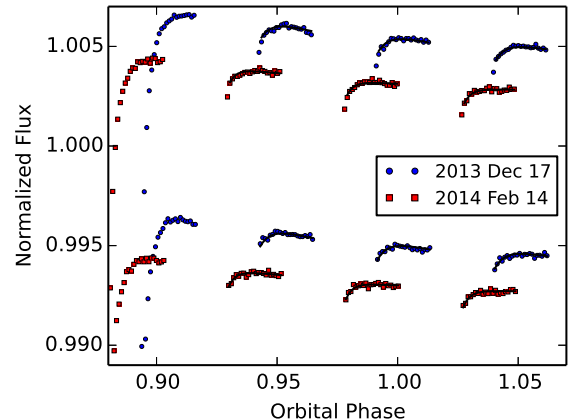}
\caption{\label{fig:rawlc}{
{\em HST}/WFC3 white light curves of GJ~436.  Blue circles depict the 2013 Dec 17 dataset and red squares depict the 2014 Feb 14 dataset.  Typical per point uncertainties are the size of the symbols (67 ppm).  For comparison, we include sample models (black curves) that fit orbits 2 -- 4 and exclude the first point from each orbit.  The forward and reverse scan directions produce the observed 1\% flux offset within each dataset.
}}
\end{figure}

We see the same ramp systematic in our data (Figure \ref{fig:rawlc}) as observed by \citet{Kreidberg2014}, so we fit orbits 2 -- 4 using a rising exponential plus linear function of the form:
\begin{equation}
\label{eqnheq}
\Gamma(\phi) = 1 - e\sp{-r\sb{0}\phi + r\sb{1}} + r\sb{2}\phi,
\end{equation}
where $r\sb{0}$ -- $r\sb{2}$ are free parameters and the orbital phase of {\em HST} is $\phi = (\Delta t+0.02)$ modulo {\em HST's} orbital period of 95.7 minutes ($\Delta t$ is the time in days since the visit's first exposure).  This model component has common parameter values to all modeled {\em HST} orbits within a visit.  In addition to not using the first orbit, we do not model the first measurement within each orbit because these points are systematically low relative to the best-fit models.  

Along with an exponential ramp dependence, many teams report visit-long trends in the light curves \citep[e.g., ][]{Berta2012,Swain2012,Sing2013,Knutson2014,Stevenson2014b}.  Most groups only explore linear functions similar in form to:
\begin{equation}
\label{eqnlin}
R(t) = 1 + r\sb{3}(t-t\sb{0});
\end{equation}
however, in previous publications \citep{Stevenson2014a,Stevenson2014c} we report achieving better fits using quadratic functions of the form:
\begin{equation}
\label{eqnquad}
R(t) = 1 + r\sb{3}(t-t\sb{0}) + r\sb{4}(t-t\sb{0})\sp{2}.
\end{equation}
In both cases, $t-t\sb{0}$ is the time relative to the expected transit midpoint and $r\sb{3}$ and $r\sb{4}$ are free parameters.  The full light curve model takes the form:
\begin{equation}
\label{eqnfull}
F(t,\phi) = F\sb{S}T(t)\Gamma(\phi)R(t),
\end{equation}
where $F(t,\phi)$ is the measured flux at time $t$ and {\em HST} orbital phase $\phi$; F\sb{s} is the scan-direction-dependent system flux outside of transit (see Figure \ref{fig:rawlc}); and $T(t)$ is the primary transit model component using the uniform-source equations from \citet{mandel2002} (i.e., no limb darkening).  $\Gamma(\phi)$ and $R(t)$ are the systematic model components defined by Equations \ref{eqnheq} -- \ref{eqnquad}.  

We perform a joint, simultaneous fit of both WFC3 observations and share common transit-model parameter values.  Since we cannot observe the entire transit within a single HST orbit, the semi-major axis and inclination are unconstrained; therefore, we fix these parameters to their published values \citep[$a/R\sb{\star}$ = 9.10, $i$ = 85.17$^{\circ}$, ][]{Stevenson2012}.  We estimate uncertainties using differential-evolution Markov-Chain Monte Carlo \citep[DE-MCMC,][]{terBraak2006,terBraak2008} and generate plots of RMS versus bin size to assess the significance of time-correlated noise in the data \citep{Pont2006,Winn2008b}.  We determine the noise to be white; therefore, there is no need to inflate uncertainty estimates.

%%%%%%%%%%%%%%%%%%%%%%%%%%%%%%%%%%%%%%%%%%%%%%%%%%%%%%%%%%%%%%%%%%%%%%%%%%%%%%%
\section{RESULTS AND DISCUSSION}
\label{sec:results}
%%%%%%%%%%%%%%%%%%%%%%%%%%%%%%%%%%%%%%%%%%%%%%%%%%%%%%%%%%%%%%%%%%%%%%%%%%%%%%%

\subsection{Search for UCF-1.01}

\begin{figure*}[tb]
\centering
\includegraphics[width=0.9\linewidth,clip]{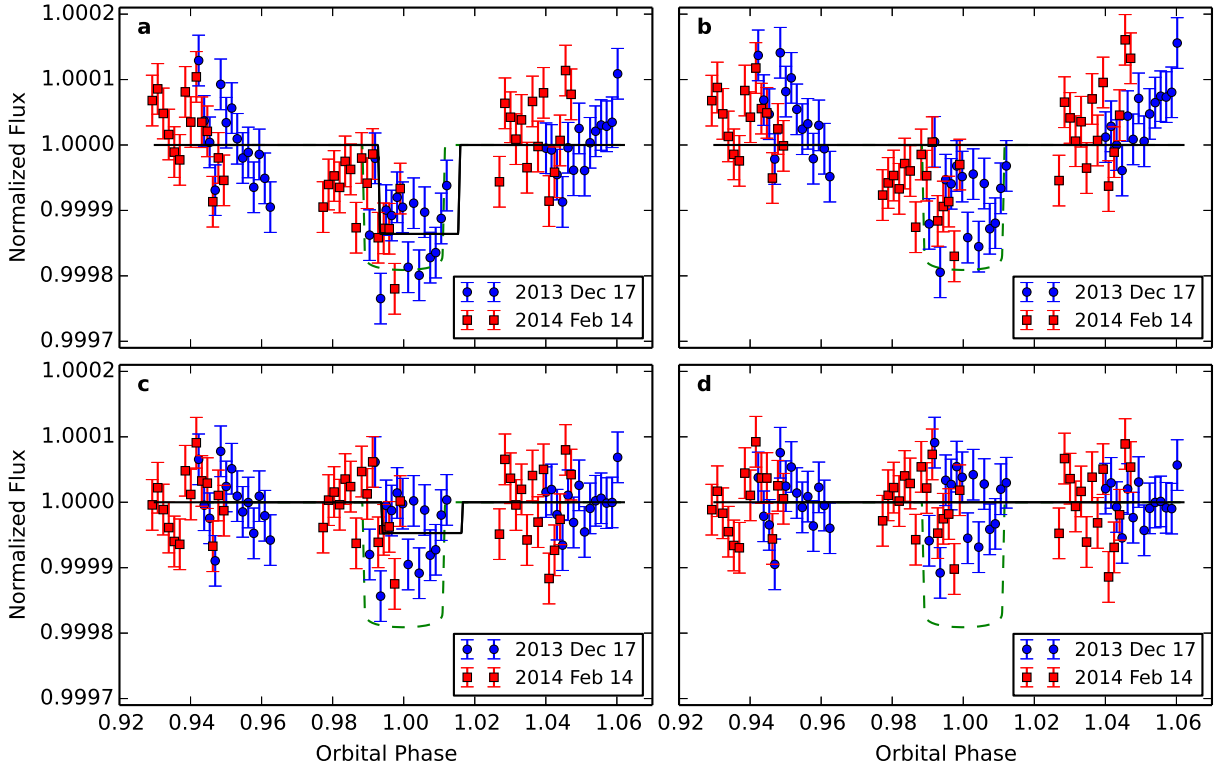}
\caption{\label{fig:lc}{
Binned {\em HST}/WFC3 white light curves of GJ~436 using four different model fits.  The systematics-removed data (blue circles and red squares) are normalized to the system flux, binned for clarity, and use 1$\sigma$ uncertainties.  Panels {\bf a} and {\bf b} depict best fits using a linear trend; panels {\bf c} and {\bf d} use a quadratic trend.  Panels {\bf a} and {\bf c} include a transit model; panels {\bf b} and {\bf d} do not.  The black solid line represents the best fit for each model combination.  This can be compared to the predicted transit signal (green dashed line) based on parameters derived from the {\em Spitzer} observations.  The bottom two panels achieve comparable BIC values (see Table \ref{tab:Compare1}) and are more likely solutions than the top two panels.
}}
\end{figure*}

In this section, we explore various model fits to the white light curves to determine the probability that we detect a sub-Earth-sized exoplanet at or near the predicted transit time.  When fitting orbits 2 -- 4 from each visit, we test visit-long linear and quadratic trends, each with and without a transit model.  In the case of a transit model with a quadratic trend, we apply a common $r\sb{4}$ value to both datasets because the individual values are statistically equivalent and the Bayesian Information Criterion (BIC) value favors a shared value.  From the results in Figure \ref{fig:lc} and Tables \ref{tab:Compare1} and \ref{tab:sysparams1}, we make several remarks.

\begin{table}[tb]
\centering
\caption{\label{tab:Compare1} 
Visit-Long Model Comparison From This Program}
\begin{tabular}{ccccc}
    \hline
    \hline
    Panel\tablenotemark{a}       
                & Model       & $R$\sb{p}/$R\sb{\star}$   & SDNR\tablenotemark{b} & $\Delta$BIC   \\
                &             &                           & (ppm)                 &               \\
    \hline
    a           & Linear      & 0.0117 {\pm} 0.0005       & 93.6, 86.3            & 62.6          \\
    b           & Linear      & --                        & 109.7, 96.9           & 201.3         \\
    c           & Quadratic   & 0.0069 {\pm} 0.0020       & 87.3, 78.9            & 0.0           \\
    d           & Quadratic   & --                        & 86.7, 81.0            & 1.2           \\
    \hline
\end{tabular}
\tablenotetext{1}{Letters coincide with panels in Figure \ref{fig:lc}.}
\tablenotetext{2}{SDNR values are listed in order of 2013 Dec 17 then 2014 Feb 14.}
\end{table}

\begin{table*}[tb]
\centering
\caption{\label{tab:sysparams1} 
Systematic Model Parameters From This Program}
\begin{tabular}{ccccccc}
    \hline
    \hline
    Label   & Panel\tablenotemark{a}       
                    & r\sb{0}       & r\sb{1}           & r\sb{2}               & r\sb{3}                & r\sb{4}              \\
    \hline
    2013 Dec 17 & a & 240 {\pm} 30  & -2.1  {\pm} 0.5   & -0.015  {\pm} 0.005   & -0.01040 {\pm} 0.00016 & --                   \\
    2013 Dec 17 & b & 220 {\pm} 30  & -2.5  {\pm} 0.5   & -0.016  {\pm} 0.005   & -0.01040 {\pm} 0.00014 & --                   \\
    2013 Dec 17 & c & 220 {\pm} 20  & -2.4  {\pm} 0.4   & -0.017  {\pm} 0.004   & -0.01060 {\pm} 0.00015 & 0.040 {\pm} 0.006    \\
    2013 Dec 17 & d & 210 {\pm} 20  & -2.5  {\pm} 0.4   & -0.017  {\pm} 0.004   & -0.01068 {\pm} 0.00015 & 0.055 {\pm} 0.005    \\
    \hline
    2014 Feb 14 & a & 360 {\pm} 50  & -0.1  {\pm} 0.9   & 0.007  {\pm} 0.003    & -0.00942 {\pm} 0.00015 & --                   \\
    2014 Feb 14 & b & 330 {\pm} 40  & -0.4  {\pm} 0.7   & 0.001  {\pm} 0.002    & -0.00943 {\pm} 0.00015 & --                   \\
    2014 Feb 14 & c & 340 {\pm} 40  & -0.3  {\pm} 0.7   & 0.003  {\pm} 0.003    & -0.00859 {\pm} 0.00018 & 0.040 {\pm} 0.006    \\
    2014 Feb 14 & d & 330 {\pm} 40  & -0.5  {\pm} 0.7   & 0.001  {\pm} 0.003    & -0.00851 {\pm} 0.00018 & 0.043 {\pm} 0.005    \\
    \hline
\end{tabular}
\tablenotetext{1}{Letters coincide with panels in Figure \ref{fig:lc}.}
\end{table*}

First, using a linear function, we detect a transit signal with a shared midpoint that is 6.1 {\pm} 3.1 minutes later than predicted assuming no TTVs.  The various TTV solutions described in Section \ref{sec:ttvs} predict transit midpoints 6 -- 13 minutes earlier than the measured time; however, the uncertainty on these solutions is {\pm}8 minutes.  We find a best-fit radius ratio that is 2.0$\sigma$ below the value reported by \citet[$R$\sb{p}/$R\sb{\star}$ = 0.0138 {\pm} 0.0009]{Stevenson2012}.  The significance of this detection hinges on the decrease in flux at the end of the third orbit of the 2014 Feb 14 visit (see Figure \ref{fig:lc}).  Despite the fact that the third orbit of the first visit is significantly lower than the linear trend, these data do not offer any additional evidence of a transit detection.  This is because the third orbit could either rest entirely within the transit (no ingress/egress) or the linear trend could be insufficient in modeling the visit-long systematic.

Second, regardless of the presence of a transiting planet, a visit-long quadratic trend provides a better fit (>\ttt{13} times more probable) than a linear trend to the available data.  This point is important because analyses that do not consider a quadratic trend when fitting the visit-long systematic may be reporting incorrect absolute transit and/or eclipse depths.  As discussed in Section \ref{sec:gj436b}, this may also explain reported inconsistencies in the measured white light curve depths over repeated visits.

Third, a transit detection from a sub-Earth-sized exoplanet is indeterminate with a quadratic trend.  The difference in BIC values for scenarios with and without a transit model component is 1.2 (see Table \ref{tab:Compare1}).  Additionally, the favored radius ratio with a quadratic trend is 0.0069 {\pm} 0.0020, which is inconsistent with previous findings by 3.1$\sigma$ \citep{Stevenson2012}.

\subsection{Injecting and Recovering a Fake Transit Signal}

\begin{figure}[tb]
\centering
\includegraphics[width=0.9\linewidth,clip]{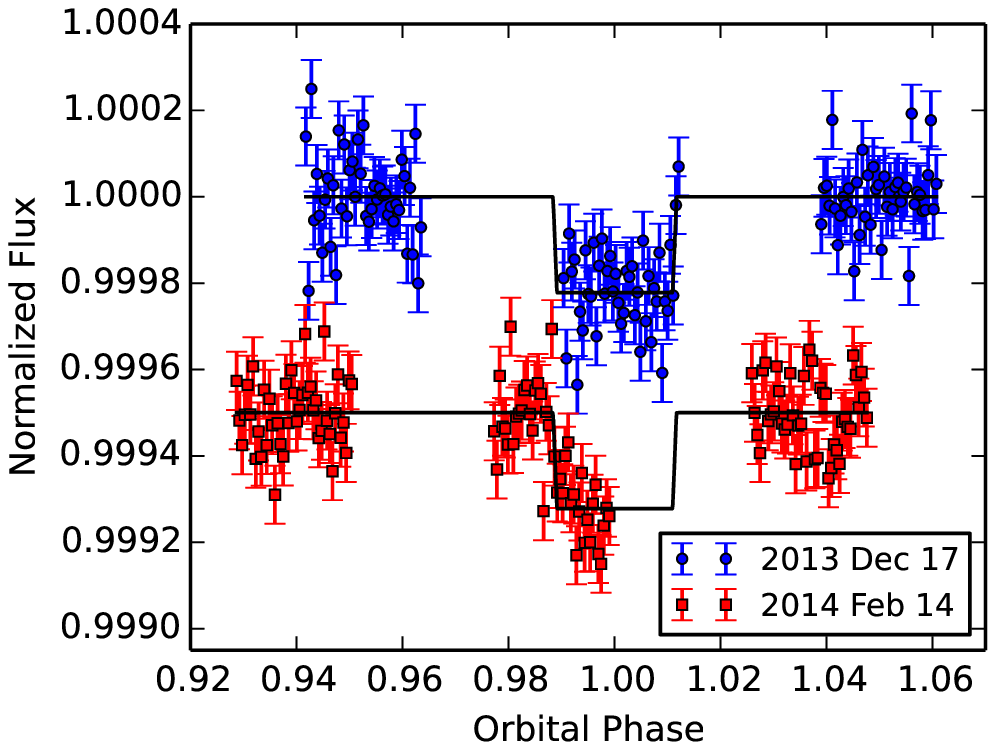}
\caption{\label{fig:fakeTr}{
{\em HST}/WFC3 white light curves of GJ~436 with an injected transit signal.  Using the predicted transit depth and midpoint for UCF-1.01, we inject a fake transit signal into the 2013 Dec 17 (blue circles) and 2014 Feb 14 (red squares) datasets.  The data points display 1$\sigma$ uncertainties and are offset vertically for clarity.  The best-fit models (black curves) recover the injected signal at the correct transit midpoint and depth.
}}
\end{figure}

To test our ability to recover the expected transit signal within the WFC3 data, we inject a fake transit signal into our white light curves using the predicted transit depth and midpoint for UCF-1.01.  Fitting the visit-long systematic with a quadratic trend, we run our DE-MCMC routine and recover the injected signal (see Figure \ref{fig:fakeTr}).  Relative to the injected values, the measured transit midpoint is 6 seconds (0.4$\sigma$) later and the measured transit depth (222 {\pm} 15 ppm) is 2.1$\sigma$ deeper.  In the case of the injected transit, we find that a fit that includes a transit model component is $\sim$\ttt{41} times more probable than a fit without a transit model component.  We conclude that the WFC3 data are sensitive to transits from a sub-Earth-sized exoplanet in the GJ 436 system.

\subsection{Reanalysis of Previously-Published WFC3 data}
\label{sec:gj436b}

To determine the necessity of using a quadratic function to model visit-long, time-dependent systematics in other WFC3 datasets, we reanalyze four additional observations of GJ 436 using the G141 grism (GO Program 11622, PI: Heather Knutson).  {\em HST} acquired these data using only the forward scan direction at a rate of 0.99{\arcsec}s\sp{-1}.  This produced a typical peak pixel value of 38,000 electrons.  \citet{Knutson2014} provide additional details of the observations, which acquired transits of GJ 436b.  We use the same reduction and white light-curve fitting routines as those described in Section \ref{sec:obs}.  We apply a 90 pixel long extraction box and perform background subtraction using the median value of each column.  In our joint fit, we share common $a/R\sb{\star}$, $\cos i$, and linear limb darkening parameter values.

\begin{figure}[tb]
\centering
\includegraphics[width=0.9\linewidth,clip]{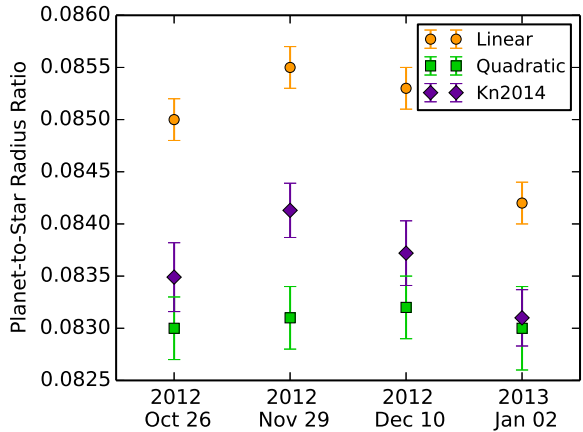}
\caption{\label{fig:rprs}{
Comparison of measured GJ 436b planet-to-star radius ratios.  Using a linear trend, we achieve inconsistent $R$\sb{p}/$R\sb{\star}$ values over the four visits.  A quadratic trend achieves a better fit and produces consistent $R$\sb{p}/$R\sb{\star}$ values.  For comparison, we also include the results published by \citet{Knutson2014}.  They used a linear trend to fit the visit-long systematic.
}}
\end{figure}

We test both linear and quadratic functions when fitting orbits 2 -- 4 of the white light curves and display the results in Table \ref{tab:Compare} and Figure \ref{fig:rprs}.  Using a linear trend, we find that the models do not achieve a good fit to the available data ($\chi^2_{\nu}$ = 1.29 -- 2.14, similar to \citet{Knutson2014}) and the planet-to-star radius ratios ($R$\sb{p}/$R\sb{\star}$) are inconsistent by up to 4.6$\sigma$.  Although our radius ratios are systematically higher than those of \citet{Knutson2014}, they have similar relative values.  This offset may be the result of finding different best-fit $a/R\sb{\star}$ and $i$ values.  For comparison, \citet{Knutson2014} fit orbits 2 -- 4 using a linear function to model the visit-long systematic and the same exponential plus linear function to model the orbit-long systematic.

Using a quadratic trend, we achieve a typical $\chi^2_{\nu}$ value of 1.06 and a smaller BIC value compared to the linear trend ($\Delta$BIC = 209.3).  Furthermore, the measured radius ratios are mutually consistent at the 1$\sigma$ level.  Using a common $R$\sb{p}/$R\sb{\star}$ for all four transits, we list best-fit parameters in Table~\ref{tab:params}.  We conclude that a quadratic function is the appropriate choice for these white light curves when modeling visit-long, time-dependent systematics.  Looking at the range of best-fit values for $r\sb{3}$ and $r\sb{4}$, we see that they are fairly consistent between datasets but not currently predictable for future analyses.

In addition to the white light curve reanalysis of GJ 436b, we repeated the spectroscopic analysis performed by \citet{Knutson2014}.  We also find the planet's transmission spectrum to be featureless and inconsistent with a cloud-free, hydrogen-dominated atmosphere.  Additionally, when we fix the remaining transit and limb-darkening parameters to their best-fit values, we compute comparable transit depth uncertainties to those reported by \citet{Knutson2014}.

\begin{table}[tb]
\centering
\caption{\label{tab:Compare} 
Visit-Long Model Comparison From Program 11622}
\begin{tabular}{ccccc}
    \hline
    \hline
    Obs. Date   & Model     & $R$\sb{p}/$R\sb{\star}$\tablenotemark{a}  
                                                    & SDNR  & $\chi^2_{\nu}$    \\
                &           &                       & (ppm) &                   \\
    \hline
    2012 Oct 26 & Linear    & 0.0850 {\pm} 0.0002   & 87.2  & 1.86              \\
    2012 Nov 29 & Linear    & 0.0855 {\pm} 0.0002   & 91.8  & 1.99              \\
    2012 Dec 10 & Linear    & 0.0853 {\pm} 0.0002   & 95.4  & 2.14              \\
    2013 Jan 02 & Linear    & 0.0842 {\pm} 0.0002   & 74.1  & 1.29              \\
    \hline
    2012 Oct 26 & Quadratic & 0.0830 {\pm} 0.0003   & 61.7  & 0.94             \\
    2012 Nov 29 & Quadratic & 0.0831 {\pm} 0.0003   & 68.3  & 1.12             \\
    2012 Dec 10 & Quadratic & 0.0832 {\pm} 0.0003   & 65.9  & 1.03             \\
    2013 Jan 02 & Quadratic & 0.0830 {\pm} 0.0004   & 70.2  & 1.17             \\
    \hline
\end{tabular}
\tablenotetext{1}{Due to correlations between the quadratic term and $R$\sb{p}/$R\sb{\star}$, the latter's uncertainties are larger than when using models with only a linear model component.  Nonetheless, the values from our quadratic fit are likely to be more accurate.}
\end{table}

\begin{table}[h]
\centering
\caption{\label{tab:params} 
Best-Fit GJ 436b Transit and Systematic Model Parameters}
\begin{tabular}{cc}
    \hline
    \hline
    Parameter                   & Value                     \\
    \hline
    $R\sb{p}/R\sb{\star}$       & 0.0831 {\pm} 0.0002       \\
    $a/R\sb{\star}$             & 13.89  {\pm} 0.08         \\ 
    $i$ [$\sp{\circ}$]          & 86.51  {\pm} 0.03         \\
    Transit Times [MJD]\tablenotemark{a}   
                                & 6226.69124 {\pm} 0.00005  \\
                                & 6261.06186 {\pm} 0.00005  \\
                                & 6271.63755 {\pm} 0.00005  \\
                                & 6295.43258 {\pm} 0.00004  \\
    $r\sb{0}$                   & 230 {\pm} 30              \\
                                & 220 {\pm} 40              \\
                                & 270 {\pm} 40              \\
                                & 170 {\pm} 30              \\
    $r\sb{1}$                   & -2.5 {\pm} 0.7            \\
                                & -2.6 {\pm} 0.7            \\
                                & -1.8 {\pm} 0.8            \\
                                & -3.7 {\pm} 0.5            \\
    $r\sb{2}$                   & -0.013  {\pm} 0.002       \\
                                & -0.011  {\pm} 0.002       \\
                                & -0.007  {\pm} 0.002       \\
                                & -0.012  {\pm} 0.003       \\
    $r\sb{3}$                   & -0.00870 {\pm} 0.00015    \\
                                & -0.00982 {\pm} 0.00016    \\
                                & -0.00846 {\pm} 0.00014    \\
                                & -0.00813 {\pm} 0.00019    \\
    $r\sb{4}$                   & 0.065 {\pm} 0.005         \\
                                & 0.076 {\pm} 0.005         \\
                                & 0.072 {\pm} 0.005         \\
                                & 0.046 {\pm} 0.007         \\
    \hline
\end{tabular}
\tablenotetext{1}{MJD = BJD\sb{TDB} - 2,450,000.}
\end{table}

%%%%%%%%%%%%%%%%%%%%%%%%%%%%%%%%%%%%%%%%%%%%%%%%%%%%%%%%%%%%%%%%%%%%%%%%%%%%%%%
\section{CONCLUSIONS}
\label{sec:concl}
%%%%%%%%%%%%%%%%%%%%%%%%%%%%%%%%%%%%%%%%%%%%%%%%%%%%%%%%%%%%%%%%%%%%%%%%%%%%%%%

The search for and confirmation of additional short-period transiting planets in the GJ 436 system remains an ongoing process.  Using the WFC3 instrument on board the {\em Hubble Space Telescope}, we were unable to definitively detect transits of the sub-Earth-sized candidate planet UCF-1.01 using the published period and ephemeris.  Some of the uncertainty in detecting UCF-1.01 revolves around how one models the visit-long time-dependent systematics.  Using a linear trend, we measure a transit signal at the predicted time and with a depth consistent with previous measurements.  However, the transit depth diminishes significantly with the use of a quadratic trend.  Using the BIC to compare the quality of our model fits, we find that a visit-long quadratic trend provides a better fit to the available data and, when using a quadratic trend, the fits are impartial to our use of a transit model ($\Delta$BIC = 1.2).  The best-fit model favors a transit depth that is inconsistent with the hypothesized value by 3.1$\sigma$. Our study of WFC3 systematics has a direct bearing on future attempts to derive accurate absolute transit and eclipse depths.

We have used today's best instruments to follow the trail -- the extrapolated ephemeris -- of this candidate planet, but the case has gone cold.  UCF-1.01 may simply not be a real planet.  But if it is, there are several reasons why we were unable to definitively detect a transit.  The planet may be smaller than originally thought (see Figure \ref{fig:lc}, panel c), transits may have occurred during {\em HST's} occultation of Earth (due to unmodeled gravitational interactions with unconfirmed additional planets, such as UCF-1.02), or its orbit may have precessed to the point where the planet no longer transits GJ 436 from our point of view.  If a sub-Earth-sized world orbiting GJ 436 is ever to be confirmed, it will likely have to be rediscovered from long-baseline monitoring of the system.  The next best opportunity to make such an observation rests with the TESS mission \citep{TESS2014}.
\\

\acknowledgments

We thank Joseph Harrington and members of the UCF Exoplanet Measurement Research Group for their efforts during the initial discovery and for feedback on subsequent proposals.  We also appreciate the many thoughtful suggestions from the annonymous referee.  We thank contributors to SciPy, Matplotlib, and the Python Programming Language, the free and open-source community, the NASA Astrophysics Data System, and the JPL Solar System Dynamics group for software and services.  Based on observations made with the NASA/ESA Hubble Space Telescope, obtained at the Space Telescope Science Institute, which is operated by the Association of Universities for Research in Astronomy, Inc., under NASA contract NAS 5-26555. These observations are associated with program GO-13338.  K.B.S. recognizes support from the Sagan Fellowship Program, supported by NASA and administered by the NASA Exoplanet Science Institute (NExScI).  J.L.B. acknowledges support from the Alfred P.~Sloan Foundation and L.K. acknowledges support from the National Science Foundation through a Graduate Research Fellowship.
\\

\bibliography{ms}

\end{document}